\documentclass{jnmp}

\usepackage{amsmath}

\JNMPnumberwithin{equation}{section}

\newtheorem{corollary}{Corollary}[section]

\theoremstyle{definition}

\begin{document}

%
\renewcommand{\evenhead}{M Szyd{\l}owski and M Biesiada}
\renewcommand{\oddhead}{Kovalevski Exponents and Integrability Properties in Class A}

%
\thispagestyle{empty}

\FirstPageHead{9}{1}{2002}{\pageref{biesiada-firstpage}--\pageref{biesiada-lastpage}}{Letter}

\copyrightnote{2002}{M Szyd{\l}owski and M Biesiada}

\Name{Kovalevski Exponents\\
 and Integrability Properties in Class A\\
Homogeneous Cosmological Models}
\label{biesiada-firstpage}

\Author{Marek SZYD{\L}OWSKI~$^\dag$ and Marek BIESIADA~$^\ddag$}

\Address{$^\dag$~Jagellonian University, Astronomical Observatory,
 Orla 171, 30-504 Cracow, Poland\\
~~E-mail: szydlo@oa.uj.edu.pl\\[10pt]
$^\dag$~Department of Astrophysics and Cosmology, University of Silesia,\\
~~Uniwersytecka 7, 40-007 Katowice, Poland \\
~~E-mail: mb@imp.sosnowiec.pl}

\Date{Received March 13, 2001; Revised August 10, 2001;
Accepted August 17, 2001}

\begin{abstract}
\noindent
Qualitative approach to homogeneous anisotropic Bianchi class A models in
terms of dynamical systems reveals a hierarchy of invariant manifolds. By
calculating the Kovalevski Exponents according to  Adler - van
Moerbecke method we discuss how algebraic integrability property is
distributed in this class of models. In particular we find that
algebraic nonintegrability of vacuum Bianchi ${\rm VII}_0$ model is inherited
by more
general Bianchi VIII and Bianchi IX vacuum types. Matter terms (cosmological
constant, dust and radiation) in the Einstein equations typically generate
irrational or complex Kovalevski exponents in class A homogeneous models thus
introducing an element of nonintegrability even though the respective vacuum
models are integrable.
\end{abstract}

\section{Introduction}

{\advance\baselineskip-0.10pt
Einstein's theory of General Relativity provides a very elegant geometric
picture of gravitational interaction.
However, such a formulation  becomes possible at the
expense of making the field equations strongly nonlinear \cite{MTW}.
Even in rather
symmetric case of spatially homogeneous space-times the dynamics of their
general types (mixmaster models) turns out to be chaotic \cite{BKL}.

On the other hand, growing importance of non-linearity in fundamental as well
as in phenomenological physical laws and richness of structure induced by
non-linear phenomena motivated strong interest in techniques allowing one to
detect and properly describe the complexity inherent to non-linear systems.
One such approach aimed at investigating the integrability of non-linear
systems has been initiated in works of Kovalevski \cite{Sonia} who in her
famous contribution to the problem
of rotating rigid body considered the
behaviour of the solution near the essential singularities in the complex plane.
Similar intuition that the position of singular points (essential singularities ) depends on (arbitrary)
integration constants and consequently that their mobility precludes the
construction of first integrals was formulated and developed by Painlev{\'e}
\cite{Ramani}.
For the review of singularity analysis, Painlev{\'e} property and integrability
of nonlinear systems see \cite{Ramani}.

Yoshida \cite{KE} developed precise methods of
investigating the restricted case of algebraic integrability of homogeneous
dynamical systems. In particular he obtained a necessary condition for
algebraic integrability in terms of Kovalevski exponents (i.e. the nonexistence
of irrational or complex KEs).
Kovalevski Exponents \cite{Sonia,KE} allow for reconstructing a set of essential
singularities for nonintegrable systems without need of constructing the
solution.
The usefulness of calculating the Kovalevski exponents has also been emphasized
by Adler and van Moerbecke \cite{Adler}. They gave the criteria for algebraic
complete integrability for Euclidean Toda-like systems.
Pavlov \cite{Pavlov} essentially extended Adler-van Moerbecke formula into the
form suitable for studying indefinite Toda lattices. Thus from the existence of
irrational or
complex KEs we shall deduce that the system does not have additional algebraic
first integrals (i.e. besides the energy integral). Of course the nonexistence
of such first integrals does not imply nonintegrability (see e.g.
\cite{Ramani}).
On the other hand when the KEs
are integer numbers we not only know that corresponding first integral exists
but we are able to construct its form.

In this paper we calculate the Kovalevski Exponents (KEs) in the wide class of
homogeneous Bianchi spacetimes represented as hamiltonian systems with
indefinite kinetic energy form and exponential potential term. Our technique is
a generalization of the methods previously applied in studies of Euclidean Toda
systems \cite{Kozlov}.
In the present paper we applied this method for the wide class
of Bianchi models with matter and cosmological constant. Our investigations are
complementary to those of Demaret and Sheen \cite{Demaret} who performed the
Painlev{\'e} perturbative test for the Bianchi Type IX model with
perfect fluid, radiation and cosmological constant in order to predict some
probable chaotic regimes. In all cases but the cosmological constant case, the
models studied do not pass the Painlev{\'e} test, exhibit infinitely many
movable
logarithmic singularities and are therefore probably chaotic. In \cite{Latifi}
(see also \cite{Contopoulos}) it was shown that the perturbation of an exact
solution exhibits a movable transcendental essential singularity thus proving
the nonintegrability. Further numerical evidence of the nonintegrability of the
Mixmaster model by demonstrating the existence of complicated dense essential singularity
patterns and infinitely-sheeted solutions with sensitive dependence on initial
conditions can be found in the paper by Bountis and Drossos \cite{BD}.

It is known that there exists a hierarchy in the class of homogeneous
cosmological models dictated by the dimensionality of isometry groups acting on
constant time hypersurfaces \cite{Bogoyavlenskii,Hawking}. This was in fact the
strongest argument supporting the claim that mixmaster models represent a
generic dynamical regime near the cosmological singularity \cite{BKL,Hawking}.
We shall explore in this paper how this structure is reflected in the
property of algebraic integrability in the class of homogeneous anisotropic
cosmological models. }


\section{Kovalevski Exponents in brief}

In this section we shall recall the main facts concerning the Kovalevski Exponents.
For this purpose,  let us consider a system of first order ordinary differential equations:
\begin{equation} \label{ode}
\frac{d x_i}{dt} = F_i(x_1,\ldots,x_n),
\end{equation}
where the right hand sides $F_i$ are rational functions of $x_1, \ldots ,x_n$.
Moreover, assume that the functions $F_i$ are weighted homogeneous with weights $M_i$ i.e.
\begin{equation*} 
F_i(\alpha^{g_1} x_1,\ldots,\alpha^{g_n} x_n) = \alpha^{M_i} F_i(x_1 , \ldots, x_n),
\end{equation*}
where $M_i = g_i +1$. Then the system (\ref{ode}) is similarity
invariant and there exists a particular solution
\begin{equation} \label{particular}
x_1 = c_1 t^{-g_1}, \ \ldots, \ x_n = c_n t^{-g_n},
\end{equation}
where the constants $c_i$ satisfy algebraic equations:
\begin{equation} \label{algebraic}
F_i(c_1, \ldots ,c_n) = - g_i c_i, \qquad i = 1,\ldots,n.
\end{equation}
When we consider variational equations of the
initial system (\ref{ode}) about the reference solution (\ref{particular})
and substitute $x_i = (c_i + \zeta_i) t^{-g_i}$ we obtain the linearized system
\begin{equation*} 
t \frac{d \zeta_i}{dt} = \sum_{j=1}^{n} K_{ij} \zeta_i,
\end{equation*}
where
\begin{equation*} 
K_{ij} = \frac{\partial F_i}{\partial x_j}(c_1,\ldots,c_n) + \delta_{ij} g_i.
\end{equation*}
The eigenvalues $\rho_1, \ldots ,\rho_n$ of the $K_{ij}$ matrix are called the
{\em Kovalevski Exponents.}  In \cite{KE} it has been proven that a {\em
necessary condition} for a similarity invariant system (\ref{ode}) to be
algebraically integrable is that {\em all} Kovalevski Exponents be {\em
rational}
numbers. In other words, if at least one Kovalevski Exponent is irrational or
imaginary then the system~(\ref{ode}) does not have algebraic first integrals.
This result is central to our subsequent investigation.

\section{Bianchi class A models as indefinite Toda lattices}

Einstein field equations for
homogeneous Bianchi class A models as well as for diagonal class B models
can be cast into a hamiltonian form \cite{Bogoyavlenskii}
\begin{equation}\label{hamiltonian}
{\cal H} = \frac{1}{2} \left( - p_{\alpha}^2 + p_{\beta_{-}}^2 + p_{\beta_{+}}^2\right) +
\exp{(4 \alpha)} V(\beta_{+}, \beta_{-}),
\end{equation}
where
\begin{gather*}
 V(\beta_{+}, \beta_{-}) = n_3^2 \exp{(-8 \beta_{+})} + n_1^2 \exp{(4
 \beta_{+} +  4\sqrt{3}\beta_{-})} +{} \nonumber  \\
\qquad {} + n_2^2 \exp{(4\beta_{+} - 4\sqrt{3}\beta_{-})} -
2 n_1 n_2 \exp{(4 \beta_{+})} - {}\nonumber  \\
\qquad {} -2 n_1 n_3 \exp{(-2\beta_{+} + 2
 \sqrt{3}\beta_{-})} - 2 n_2 n_3 \exp{(-2\beta_{+} - 2\sqrt{3}\beta_{-})}.
 \nonumber
 \end{gather*}
The variables $\alpha$, $\beta_{+}$, $\beta_{-}$ have the meaning of expansion
factor and anisotropy parameters respectively \cite{MTW} whereas $n_1$, $n_2$,
$n_3$ parametrize the structure constants of isometry groups (defining the type
of the spacetime).

Hamiltonian (\ref{hamiltonian}) is a {\em generalized indefinite Toda
lattice} i.e. it is of the type
\begin{equation}\label{toda}
{\cal H} = \frac{1}{2} \langle p,p \rangle + \sum_{i=1}^N c_i \exp{(a_i,q)},
\end{equation}
where $\langle \cdot\,, \, \cdot\, \rangle$ is a {\em Lorenzian} scalar product and
$(\cdot\, , \, \cdot\,)$ is an Euclidean scalar product, ${\vec a}$ is a real vector.

\section{Kovalevski Exponents for indefinite Toda Lattices}

Given generalized Toda lattice let us define new variables
\begin{equation*}
u_i = \langle a_i, p \rangle ,\qquad v_i = \exp{(a_i,q)}.
\end{equation*}
   Transformation between $(a,p)$ and $(u,v)$ variables is a generalization of
Flaska transformation.

Then one obtains an autonomous dynamical system with polynomial right hand
sides
\begin{gather} \label{autonomous}
\dot v_i = u_i v_i, \qquad
\dot u_i = \sum^N_{j=1} M_{ij} v_{j}, \qquad
M_{ij} = - c_j \langle  a_i, a_j \rangle.
\end{gather}
The system (\ref{autonomous}) is of the type (\ref{ode})
and one can easily check that it is similarity invariant and possesses
a particular solution (see (\ref{particular})):
\begin{equation*}
u_i = \frac{U_i}{t}, \qquad v_i = \frac{V_i}{t^2},
\end{equation*}
where the constants $U_i$, $V_i$ fulfill the following algebraic
equations (see (\ref{algebraic})):
\begin{equation}\label{constants}
- 2 V_i = U_i V_i, \qquad - U_i = \sum^n_{j=1} M_{ij} V_{j}.
\end{equation}

In order to discuss the single-valuedness of solutions one investigates
behavior of solutions in the vicinity of particular solutions. Let us hence
consider the first variation
\begin{equation} \label{variation}
\frac{d}{dt} \delta u_i = \sum^n_{j=1} M_{ij} \delta v_j,\qquad
\frac{d}{dt} \delta v_i = \frac{U_i \delta v_i}{t} + \frac{V_i \delta
u_i}{t^2}.
\end{equation}
We seek for the particular solutions of (\ref{variation}) in the form
\begin{equation*}
\delta u_i = \xi_i t^{\rho-1}, \qquad \delta v_i = \eta_i t^{\rho-2},
 \end{equation*}
where $\xi_i$, $\eta_i$ are constants and $\rho$ are called Kovalevski
Exponents and satisfy the following system of algebraic equations:
\begin{equation} \label{KEs}
(\rho - 2 - U_i) \eta_i = V_i \xi_i, \qquad
(\rho - 1) \xi_i = \sum^n_{j=1} M_{ij} \eta_j.
\end{equation}

The general procedure for extracting the Kovalevski Exponents is to solve
the systems (\ref{constants}) and (\ref{KEs}).

Some specific cases can clearly be distinguished:

(i) $\exists \; i: V_i \neq 0$; $\forall \; i \neq j: V_j = 0,$

(ii) $\exists \; (i,j): V_i \neq 0, V_j \neq 0$; $\forall \; k \neq i, k \neq j: V_k
=0,$

(iii) $\exists \; (i,j,k): V_i \neq 0, V_j \neq 0, V_k \neq 0$; $\forall \; l \neq i, l
\neq j, l \neq k: V_l = 0,$

\noindent
and so on until the dimension $N$ of the system is
reached. Integer Kovalevski Exponents are necessary for solution to be
meromorphic.

In the case (i) $\rho_1 = -1$ and $\rho_2 = 2$ are always solutions, other
Kovalevski Exponents can be calculated from the generalized Adler - van
Moerbecke formula:
\begin{equation} \label{adler}
\rho = 2 - \frac{M_{ij}}{M_{jj}},
\end{equation}
where $i \neq j$, $M_{jj} \neq 0$ and $j$ is such that $V_j \neq 0$.

In the case (ii) one should solve the system:
\begin{equation} \label{case2}
\left[ - M_{jj} + \frac{\rho (\rho - 1)}{V_j}\right] \left[\frac{\rho (\rho -1)}{V_i} - M_{ii}\right]
- M_{ij} M_{ji} = 0.
\end{equation}
For this purpose it is useful to build an auxillary matrix
\begin{equation*}
{\hat M}_{ij} =
\left(
\begin{array}{cc}
M_{ii}  & M_{ij}\\
M_{ji} & M_{jj}
\end{array}
\right).
\end{equation*}

\section{Vacuum Bianchi class A models}

In this class of model we have:
\begin{gather*}
\vec{a_1}=(4,-8,0),\qquad \vec{a_2} = \left(4,4,4\sqrt{3}\right), \qquad
\vec{a_3}=\left(4,4,-4\sqrt{3}\right), \nonumber \\
\vec{a_4} = (4,4,0), \qquad \vec{a_5}=\left(4,-2,2\sqrt{3}\right),
\qquad \vec{a_6} = \left(4,-2,-2\sqrt{3}\right) 
\end{gather*}
and
\begin{equation*}
\vec{c} = \left(n_1^2, n_2^2, n_3^2, -2n_1 n_2, -2n_2 n_3, -2n_1 n_3\right).
\end{equation*}
The respective Cartan matrix reads:
\begin{equation} \label{cartan}
M_{ij} = -48 \left[
\begin{array}{cccccc}
n_1^2 & -n_2^2 & -n_3^2 & 2 n_1 n_2 & 0 & 0 \\
- n_1^2 & n_2^2 & -n_3^2 & 0 & 0 & 2 n_1 n_3 \\
- n_1^2 & - n_2^2 & n_3^2 & 0 & 2 n_2 n_3 & 0 \\
- n_1^2 & 0 & 0 & 0 & n_2 n_3 & n_1 n_3 \\
0 & 0 & - n_3^2 & n_1 n_2 & 0 & n_1 n_3 \\
0 & - n_2^2 & 0 & n_1 n_2 & n_2 n_3 & 0
\end{array}
\right].
\end{equation}

Application of the generalized Adler - van Moerbecke formula (whenever there
is only one $V_i \neq 0$) shows that all Kovalevski Exponents (corresponding
to this case) are integer and equal:
\begin{equation*}
\rho_1 = -1,\qquad \rho_2= 2,\qquad \rho_3 = 4.
\end{equation*}
Let us now check if ${\hat M}_{ij}$ is degenerate. For this purpose we build
an auxillary traceless symmetric matrix,
\begin{equation} \label{hat}
\hspace*{-4mm}
[\det {\hat M}_{ij}] =\! \left[\!\!
\begin{array}{cccccc}
0 & 0 & 0 & 96 n_1^3 n_2 & 0 & 0 \\
0 & 0 & 0 & 0 & 0 & 96 n_1 n_2^2  n_3\\
0 & 0 & 0 & 0 & 96 n_2 n_3^3 & 0 \\
96 n_1^3 n_2\! & 0 & 0 & 0 & - 48 n_1 n_2^2 n_3 & \!- 48 n_1^2 n_2 n_3 \\
0 & 0 & \! 96 n_2 n_3^3\! & \!-48 n_1 n_2^2 n_3\! & 0 & \!-48 n_1 n_2 n_3^2 \\
0 & \!96 n_1 n_2^2 n_3 \! & 0 & \!-48 n_1^2 n_2 n_3\! & \!-48 n_1 n_2 n_3^2 & 0
\end{array}\!\!
\right]\!.
\end{equation}

An inspection of (\ref{hat}) allows us to formulate the following

\medskip

{\sc Conclusions}
\begin{itemize}
\itemsep0mm
\item{} for Bianchi I and Bianchi II models every block matrix is degenerate;
\item{} for Bianchi ${\rm VI}_0$ and Bianchi ${\rm VII}_0$ models there is
only one pair $(V_1, V_4)$ leading to nondegenerate matrix ${\hat M}_{14}$;
respective Kovalevski Exponents read: B(${\rm VI}_0$) $\rho_{1,2} =
\frac{1}{2} (1 \pm i\sqrt{15})$, $\rho_3 = -1$, $\rho_4 = 2$ and for
B(${\rm VII}_0$): $z = \rho(\rho -1) = 1 \pm i \sqrt{7}$;
\item{} from mixmaster models (i.e. Bianchi IX and VIII) there is six pairs
of $(V_i,V_j)$ leading to nondegenerate ${\hat M}_{ij}$; respective Kovalevski
exponents are the same for Bianchi IX and VIII models and read: $z = 1 \pm
i\sqrt{7}$ for pair $(1,4)$ and $z = \pm 2$ for pairs
$(3,5)$; $(4,5)$; $(4,6)$; $(5;6)$. Because $z = \rho(\rho -1)$ then $z= 2$ corresponds
to $\rho_1 = -1$, $\rho_2 = 2$ and $z = -2$ corresponds to $\rho_{1,2} =
\frac{1}{2}(1\pm i \sqrt{7})$.
\end{itemize}

\begin{corollary}
Bianchi type {\rm II} model is algebraically integrable.
\end{corollary}

\begin{corollary}
Mixmaster models inherit their algebraic nonintegrability from Bianchi ${\rm
VII}_0$ model.
\end{corollary}

\section{Homogeneous models with cosmological constant,\\
dust and radiative matter}

One can now consider cosmological models with nonvanishing matter terms. They
are described by the hamiltonian:
\begin{gather*}
{\cal H} = \frac{1}{2} \left(- p_{\alpha}^2 + p_{\beta_{+}}^2 + p_{\beta_{-}}^2\right) +
\exp{(4 \alpha)} V(\beta_{+},\beta_{-}) +{}\nonumber\\
\qquad {}+ \Lambda \exp{(6 \alpha)} +
\mu_{\rm dust} \exp{(3 \alpha)} + \Gamma_{\rm rad} \exp{(2 \alpha)},
\end{gather*}
where $\Lambda$ as usually  denotes the cosmological constant and the
subsequent terms denote other matter sources i.e. the amount of
energy-momentum carried by dust and radiative matter, respectively.

\subsection{Models with cosmological constant}

Let us consider the case $\Lambda \neq 0$, $\Gamma_{\rm rad} = \mu_{\rm dust} = 0.$
In addition to six
${\vec a_i}$ vectors in the decomposition (\ref{toda}) there exists now one
more vector:
\begin{equation*}
{\vec a_7} = (6,0,0)
\end{equation*}
and subsequent formulae (\ref{cartan}) and (\ref{hat}) are modified
correspondingly. From the formula (\ref{adler}) one can calculate the following
additional Kovalevski Exponents:
\begin{gather*}
\rho_i = 2 - 2 \frac{\langle {\vec a_i},{\vec a_7}\rangle}{\langle{\vec a_7},{\vec a_7}\rangle} =
\frac{2}{3} \qquad \mbox{for} \quad i=1,\ldots,6, \nonumber \\
\rho_i = 2 - 2 \frac{\langle{\vec a_7},{\vec a_i}\rangle}{\langle{\vec a_i},{\vec a_i}\rangle} = 3
 \qquad \mbox{for} \quad  i=1,2,3.
\end{gather*}
Other Kovalevski Exponents obtained from (\ref{case2}) are:
\begin{equation*}
z_{1,2} = -1 \pm \frac{{\sqrt{7}}}{2}, \qquad
z_{1,2} = -\frac{3}{2} \pm \frac{{\sqrt{7}}}{2}.
\end{equation*}
Recall that $z = \rho(\rho -1)$.
Hence we obtained irrational and complex values of the Kovalevski Exponents. Their existence
does not depend on the value of cosmological constant. This proves that the
presence of $\Lambda$ introduces an element of nonintegrability in the model.

\subsection{Models with dust matter}

Now we choose $\mu_{\rm dust} \neq 0$ and fix $\Lambda = \Gamma_{\rm rad} = 0.$
Additional seventh
${\vec a_i}$ vector in the decomposition (\ref{toda}) is:
\begin{equation*}
{\vec a_7} = (3,0,0).
\end{equation*}
Adler - van Moerbecke formula (\ref{adler}) gives
the following Kovalevski Exponents:
\begin{gather*}
\rho_i = 2 - 2 \frac{\langle{\vec a_i},{\vec a_7}\rangle}{\langle{\vec a_7},{\vec a_7}\rangle} =
\frac{2}{3} \qquad \mbox{for} \quad i=1,\ldots,6, \nonumber \\
\rho_i = 2 - 2 \frac{\langle{\vec a_7},{\vec a_i}\rangle}{\langle{\vec a_i},{\vec a_i}\rangle} =
\frac{5}{2}  \qquad \mbox{for} \quad i=1,2,3.
\end{gather*}
Further Kovalevski Exponents obtained from (\ref{case2}) read:
\begin{equation*}
z_{1,2} = -\frac{1}{20} \pm \frac{\sqrt{591}}{60}.
\end{equation*}
Therefore
irrational and complex values of the Kovalevski Exponents exist irrespectively
of concrete value of $\mu$ or Bianchi type.

\subsection{Models with radiative matter}

We take $\Gamma_{\rm rad} \neq 0$ and $\Lambda = \mu_{\rm dust} = 0$.
Additional vector in the decomposition (\ref{toda}) reads now:
\begin{equation*}
{\vec a_7} = (2,0,0).
\end{equation*}
Adler - van Moerbecke
formula (\ref{adler}) yields
\begin{gather*}
\rho_i = 2 - 2 \frac{\langle {\vec a_i},{\vec a_7}\rangle}{\langle{\vec a_7},{\vec a_7}\rangle} = 0
\qquad \mbox{for} \quad i=1,\ldots,6, \nonumber \\
\rho_i = 2 - 2 \frac{\langle{\vec a_7},{\vec a_i}\rangle}{\langle{\vec a_i},{\vec a_i}\rangle} = 18
 \qquad \mbox{for} \quad i=1,2,3.
\end{gather*}
Proceeding along the route of (\ref{case2}) reveals that for other Kovalevski
exponents $z = \rho(\rho -1)$ parameters satisfy the following equations:
\begin{gather*}
363 z^2 - 22 z - 52/3 = 0, \qquad
z^2 - 4 z - 1/192 = 0.
\end{gather*}
Both of them lead to complex Kovalevski exponents irrespectively of
$\Gamma_{\rm rad}$ or concrete Bianchi type concerned.

\section{Conclusions}

Qualitative analysis of the dynamics of homogeneous Bianchi models revealed a
remnant of the hierarchy of invariant manifolds. In the Bogoyavlenskii~\cite{Bogoyavlenskii}
formalism different Bianchi types lie on the components of the boundary of
compact manifold on which respective dynamical system is defined. This boundary
has its own boundaries thus generating a hierarchy in the class of homogeneous
cosmological models which could be put into one-to-one correspondence with
respective Bianchi type. The most general of homogeneous models i.e. the so
called mixmaster models are nonintegrable (e.g. \cite{Latifi,Contopoulos,BD}).
We have therefore asked the question whether there exits a respective hierarchy
of nonintegrability in the full class of homogeneous cosmological models. In
this paper we answered this question at the level of algebraic integrability by
showing that Bianchi ${\rm VI}_0$ and ${\rm VII}_0$ models are algebraically nonintegrable
since they have complex Kovalevski exponents. This property is propagated into
more general Bianchi types (IX and VIII) according to the structure of
invariant submanifolds. We have shown algebraic integrability of Bianchi type I
and II models.

We have also discussed the influence of matter terms in
the energy-momentum tensor on the property of algebraic integrability of
homogeneous cosmological models. It turned out that inclusion of dust,
cosmological constant or radiation introduces algebraic nonintegrability in the
system.
One can wonder whether the cosmological constant for example should
appear in the Kovalevski exponents and then give integer numbers in the
limit as it tended to zero. However it is well known that one cannot obtain solutions of
the Einstein equations without cosmological constant
from the respective ones with cosmological constant by taking the limit $\Lambda \to 0$.
This property is reflected also at the level of Kovalevski exponents.

The literature concerning various aspects of homogeneous cosmological models
is very big and diverse, e.g. Kramer et al.~\cite{Kramer} gives a compendium
of exact solutions of the Einstein's equations (and in large part is devoted to
homogenous models), also a dynamical systems approach initiated by
Bogoyavlenskii has considerably been developed by Wainwright and Ellis (for
the review see \cite{WE}) and chaotic aspects of Mixmaster homogeneous models
have recently been investigated (e.g. \cite{Cornish}) in more detailed way.
However, the existence of an exact solution does not imply the existence of
algebraic first integrals. Similarly, the hierarchy revealed by the dynamical
systems approach concerns the generality of solutions not the property of
their integrability. Also the works revealing dynamical complexity of
some models (in terms of fractal basin boundaries \cite{Cornish}) can by no
means be understood as a rigorous proof of their nonintegrability.
Therefore our results presented above can be perceived as complementary
to these existing in the literature and are another step toward deeper
understanding of mathematical properties of homogeneous cosmological models.

\subsection*{Acknowledgements}
Support (of M.Sz.) from the Polish State Committee for Scientific Research
Grant 2P03D 01417 is gratefully acknowledged.

\label{biesiada-lastpage}

\end{document}